\documentclass[review]{elsarticle}

\usepackage{hyperref}

\journal{Current Opinion in Biotechnology}

\begin{document}

\begin{frontmatter}

\title{Network Inference in Systems Biology: Recent Developments, Challenges, and Applications}


\author{Michael M. Saint-Antoine$^{1}$ and Abhyudai Singh$^{2}$}

\fntext[myfootnote1]{Center for Bioinformatics and Computational Biology, University of Delaware, Newark, Delaware 19716, USA}
\fntext[myfootnote2]{Electrical and Computer Engineering, University of Delaware, Newark,
Delaware 19716, USA}

\begin{abstract}
One of the most interesting, difficult, and potentially useful topics in computational biology is the inference of gene regulatory networks (GRNs) from expression data. Although researchers have been working on this topic for more than a decade and much progress has been made, it remains an unsolved problem and even the most sophisticated inference algorithms are far from perfect. In this paper, we review the latest developments in network inference, including state-of-the-art algorithms like PIDC, Phixer, and more. We also discuss unsolved computational challenges, including the optimal combination of algorithms, integration of multiple data sources, and pseudo-temporal ordering of static expression data. Lastly, we discuss some exciting applications of network inference in cancer research, and provide a list of useful software tools for researchers hoping to conduct their own network inference analyses.
\end{abstract}

\end{frontmatter}

\section{Introduction}

Networks of gene interactions, in which genes activate and repress the transcription of other genes, are responsible for much of the complexity of cellular life [1,2], and their malfunction can be disastrous for an organism [3]. So, understanding these networks has long been a goal of systems biology. Discovering gene interactions experimentally can be very difficult, and given the approximately 20,000 genes in the human genome, it is not feasible to do an experiment for every possible pair to check for an interaction. However, with traditional DNA microarray [4], or next-generation sequencing technologies, most notably (single-cell or bulk) RNA-sequencing (RNA-Seq) [5,6], one can get a quantitative peek into the transcriptomic profile of an individual cell, or a population of cells. Unfortunately, these measurement techniques involve killing the cell, so each cell can provide only one timepoint of data. A significant disadvantage of this static data is that it will not allow us to detect self-edges (in which a gene regulates itself). 

In some cases, it may be possible to construct pseudo-temporal data from static single-cell measurements. For example, a researcher may administer a stimulus to a set of cells, and then perform an RNA-Seq experiment on some cells after one hour, then on some cells after two hours, and so on, in order to collect data on the post-stimulus gene expression dynamics. In other cases, it may be possible to infer the temporal ordering of a set of gene expression measurements using computational methods (more on that in the Computational Challenges section). In this paper, the input gene expression data will be assumed to be static single cell data (meaning that each measured cell provides us with only one timepoint of data), unless otherwise noted.

\section{Problem Formulation}

   \begin{figure}[thpb]

      \centering

      \includegraphics[scale=.33]{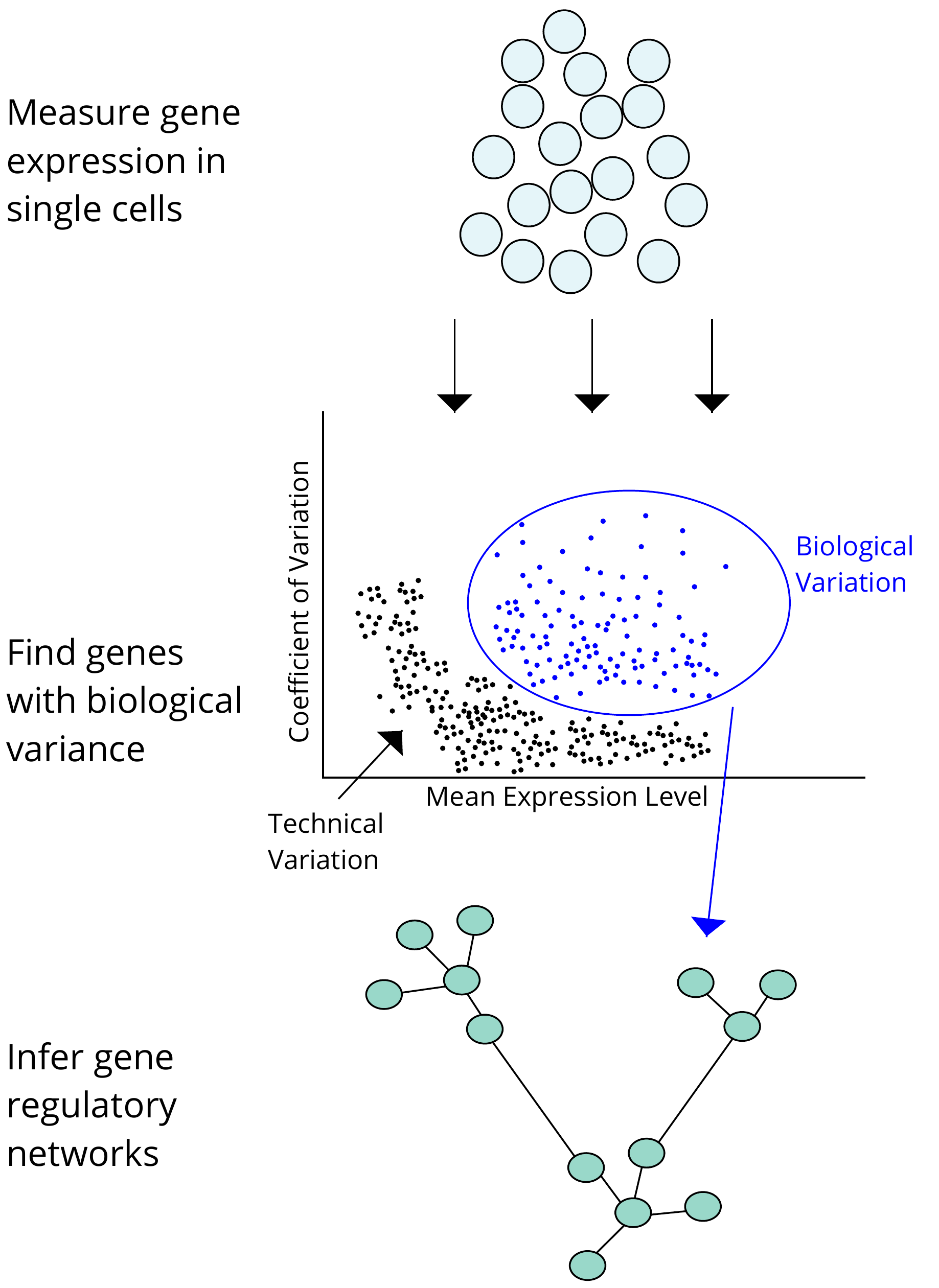}
      \caption{ A typical workflow to identity genes with high biological variance for network mapping studies.  RNA-Seq typically quantifies the expression of thousands of genes across individual cells or environmental/genetic conditions. The Coefficient of Variation or CV (Standard deviation over mean) in the expression level across cells or conditions is plotted as a function of the mean expression level for all genes. Fitting a trend line through this data provides the expected CV of a gene at given mean level, and this trend line is then used to select genes that have significantly higher CVs (representing high biological variance) while filtering out genes with low CVs that could result from technical noise.}
      \label{figurelabel}
   \end{figure}

For convenience, it can be useful to abstract the problem of network inference into a graph theory framework. Figure 1 shows the workflow of a hypothetical network inference scheme where, for example, single-cell RNA-Seq is used to measure the expression of a large number of genes in many individual cells. 
Given the large number of genes measured in these experiments it maybe prudent to select a smaller a subset of genes with high biological variance for further analysis. Such genes are typically identified by first plotting the Coefficient of Variation or CV (standard deviation over mean) in expression levels across cells with respect to the mean expression levels for all genes, and then selecting genes with significantly higher CVs than what is expected for a gene at the same mean level.  Consider $N$ genes that are identified for network analysis and let their expression levels be represented by random variables $\{X_1, X_2,...,X_N\}$. Each variable corresponds to a node in the GRN, and each edge $X_i\rightarrow X_j$ represents a regulatory relationship between $X_i$ and $X_j$. We can think of the true biological network as a real, unknown set of interactions, and our goal is to approximate it by constructing a set of weighted interactions, where each weight corresponds to our confidence that an edge exists in the true network.

A network prediction can be either directed, meaning that a prediction is made about the direction of causality for each interaction, or undirected, meaning that no such prediction is made. Directed networks are preferable, as they convey more information, but are also more difficult to construct. A network can also be signed, meaning that activation edges are labeled with a ``+'' and repression edges are labeled with a ``-'', or unsigned, meaning the edges are not labeled as positive or negative. Signed networks are important for drawing biological insights. However, unsigned networks are often used for theoretical work on network inference, since the difficult problem is determining which edges represent true interactions. Once an interaction is known, determining its sign is very easy and can be found with a simple covariance. Inferred networks have been shown to contain several motifs (smaller modules that occur more often in the network than expected by random chance), and these motifs are illustrated in Figure 2.

   \begin{figure}[thpb]

      \centering

      \includegraphics[scale=.25]{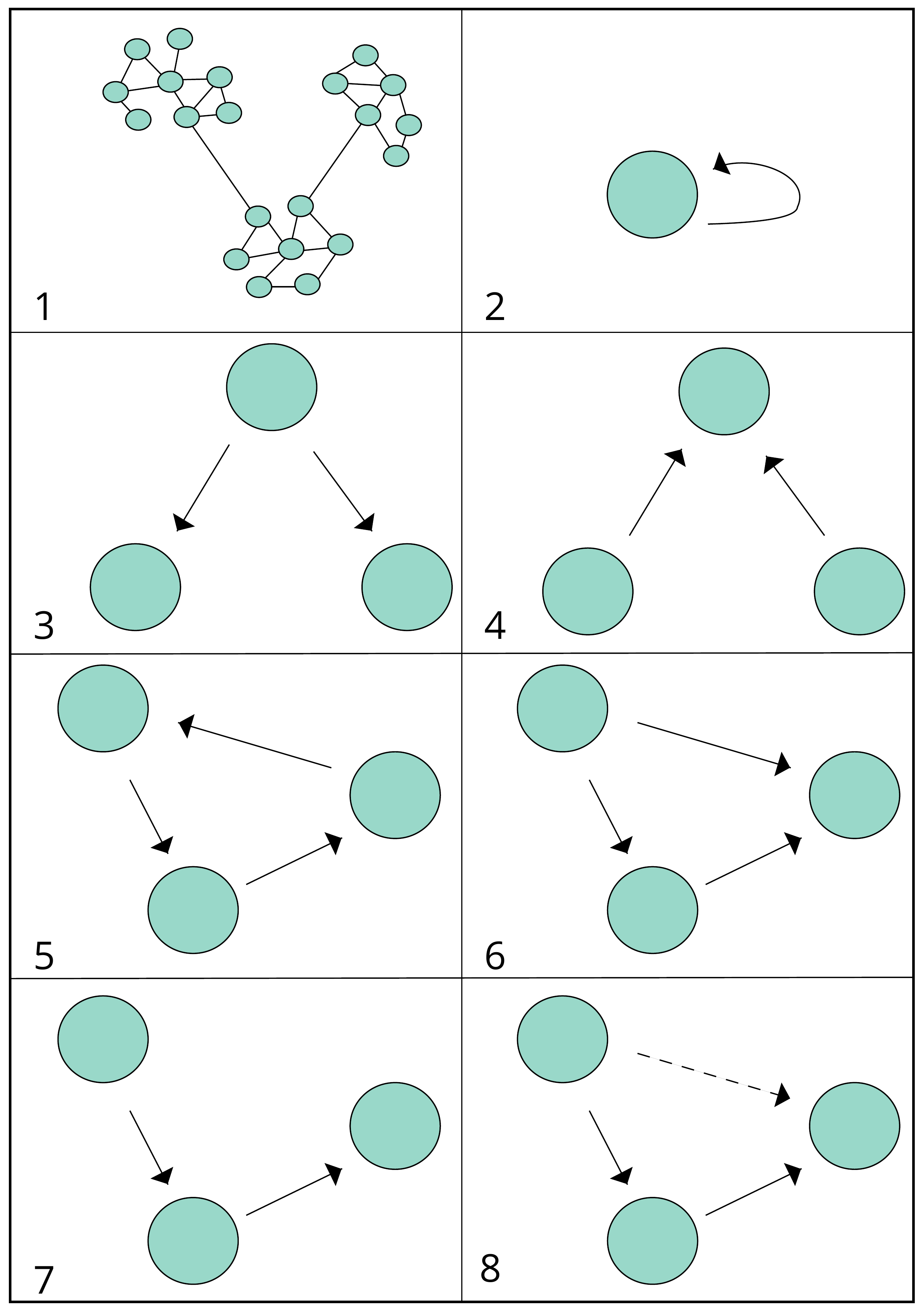}
      \caption{Illustration of some common GRN motifs. 1. Modularity (clustering of nodes), 2. Self-edge, 3. Fan-out, 4. Fan-in, 5. Feed-back loop, 6. Feed-forward loop, 7. Cascade, 8. Cascade false positive error (a feed-forward loop is incorrectly predicted). In the ``Algorithms'' section, we will discuss the strengths and weaknesses of various algorithm classes when it comes to detecting different network motifs. 
}
      \label{figurelabel}
   \end{figure}

The output of a network inference algorithm is a set of weighted edge predictions, where each edge-weight corresponds to the confidence that a real interaction exists between two genes. So, the accuracy of these algorithms can be evaluated by running them on a ``gold standard'' dataset for which the true network structure is already known. The algorithms' ability to recover the true interactions can then be scored using receiver operating characteristic (ROC) curves or precision-recall (PR) curves. Some researchers have suggested that PR curves are the superior metric [7], although both metrics are commonly used.

Fortunately for network inference researchers, several gold standard datasets are publicly available in the form of DREAM Challenges [8]. These are computational biology competitions for which researchers are invited to design new algorithms to make predictions from data, and many involve predicting GRN structures from expression data. Once each competition is over, the datasets and true network structures are released to the public, and are commonly used for algorithm evaluation [9,10,11].

\section{Algorithms}

Several broad classes of algorithms are used for network inference. An excellent introductory overview of these can be found in Huynh-Thu and Sanguinetti 2019 [12]. A thorough evaluation of the different types of algorithms can be found in Marbach et al. 2012 [9], in which many different algorithms were benchmarked on the DREAM5 gold standard networks, with varying results. To put it simply, there is no overall ``best'' class of algorithms. Rather, each has its own advantages and disadvantages, so choosing the best one depends on the context. Here, we will give a brief description of each class, but will mainly focus on recent developments and unsolved challenges. Figure 3 provides a summary of some of the properties of different algorithm types.

   \begin{figure}[thpb]

      \centering

      \includegraphics[scale=.35]{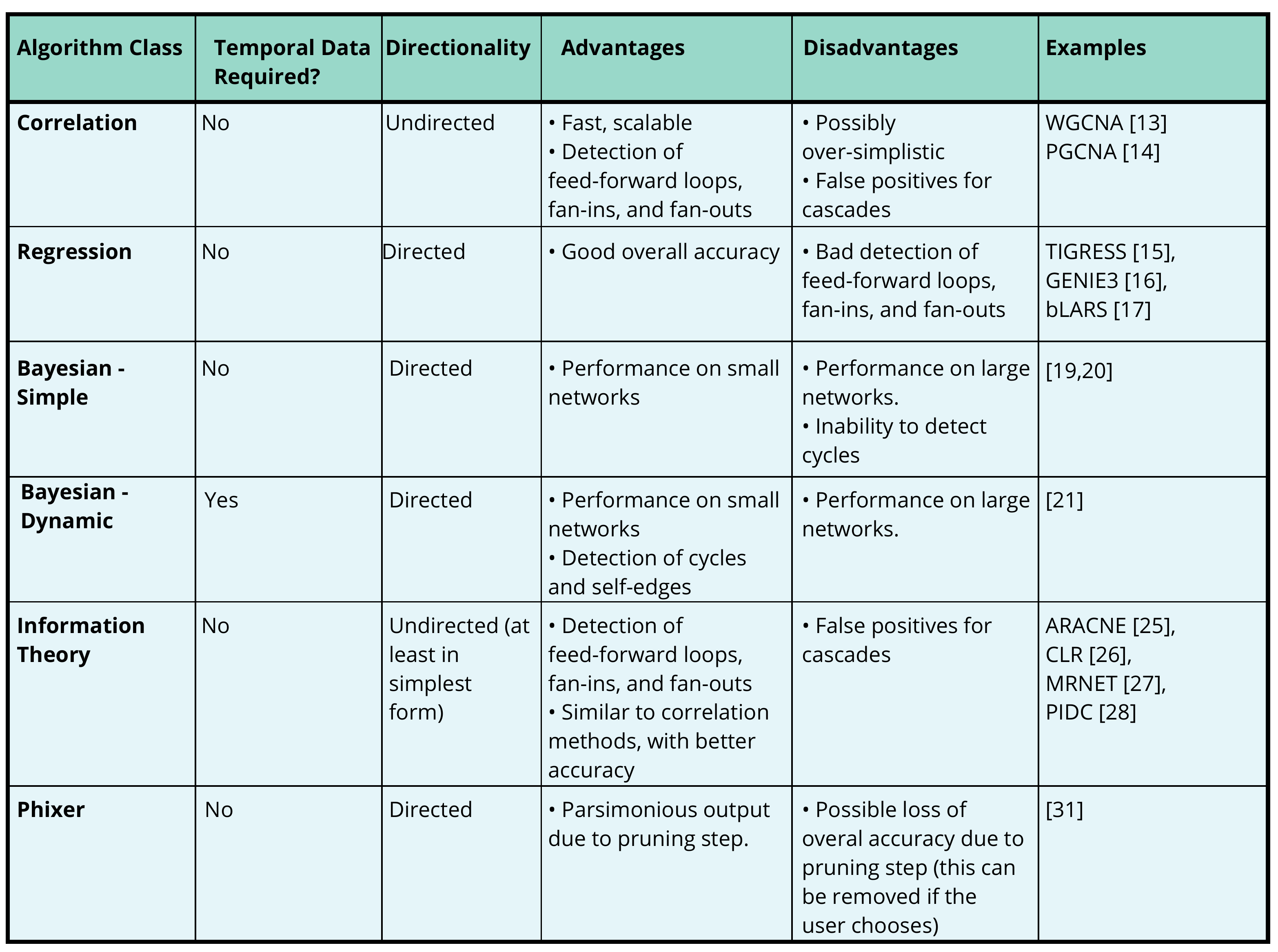}
      \caption{Summary of the properties, advantages, and disadvantages of different types of algorithms.
}
      \label{figurelabel}
   \end{figure}

\subsection{Correlation}

One of the most basic methods of network inference is to simply compute the correlation for each pair of genes. This method is very simplistic, but is also fast and scalable for large datasets. It can be especially useful in cases where researchers want a general idea of which genes are related, without caring much about causal direction or distinguishing between direct and indirect regulation. For this reason, the resulting network prediction may sometimes be referred to as a ``gene co-expression network'' rather than a GRN. In general, correlation-based algorithms tend to perform relatively well on detecting feed-forward loops, fan-ins, and fan-outs, but also have an increased false positive rate for cascades [9] (please see Figure 2 for an explanation of these motifs).

Though simplistic, correlation networks can yield powerful insights when the proper analytical tools are applied. For example, Weighted Gene Co-expression Network Analysis (WGCNA) [13], based on correlation, was an early and widely-used method in gene network analysis. Furthermore, improvements to correlation-based network inference are still being made. Care et al. 2019 [14] describes how to deal with the problem of over-connectivity in correlation networks, effectively reducing the number of edges to get to the point where the network is sparse enough for a cluster analysis to be performed.

\subsection{Regression}

Another common method of network inference is regression analysis. In its simplest form, this method involves solving the following linear regression equation to predict $X_j$ from $X_i$:

\begin{equation}
X_j= \beta_0 + \beta_1X_i + \epsilon
\end{equation}

Here, $\beta_0$ is the intercept, $\beta_1$ is the slope, and $\epsilon$ is the random error term. So, since $\beta_1$ represents the relationship between $X_j$ and $X_i$, we can assign it as the weight of the edge $X_i\rightarrow X_j$. This is only the simplest version of regression analysis, and many regression-based algorithms use more advanced methods. Two notable, highly-regarded examples are TIGRESS [15] and GENIE3 [16].

Regression methods are more computationally expensive than correlation, but also provide the advantage of predicting causal direction. In general, regression-based methods perform well compared to other methods overall, but perform poorly on the specific network motifs of feed-forward loops, fan-ins, and fan-outs [9]. It is also important to note that regression methods that resample data (by bootstrapping, for example) typically outperform regression methods that do not [9]. One limitation of regression-based methods, at least in their simplest form, is that they typically assume linear relationships between genes, and may fail to detect non-linear regulatory interactions. However, some progress has been made on this matter: bLARS [17], a modification of the Least Angle Regression method [18], is a regression-based algorithm designed to detect both linear and pre-defined non-linear interactions.

\subsection{Bayesian Methods}

Bayesian methods have long been used in gene network inference [19,20]. In these methods, an interaction between genes is represented as a conditional probability. For two genes with expression levels $X_i$ and $X_j$, the conditional probability $P (X_j | X_i)$ corresponds to the edge $X_i\rightarrow X_j$, and $X_i$ is said to be the parent of $X_j$. The graphical representation of a set of these conditional probabilities is called a Bayesian network. Given a gene expression dataset, a maximum likelihood estimation can be applied to determine which Bayesian network structure has the highest posterior probability. In other words, the goal of the maximum likelihood estimation is to determine which network structure is the most likely to have produced the observed data.

An advantage of Bayesian methods is the ease with which prior knowledge of interactions can be integrated [12]. Unfortunately, Bayesian methods are typically very computationally expensive. In general, they tend to perform poorly on large datasets compared to other methods, and may be better suited to small networks for which their heuristic searching method can more easily converge on the true optimal network structure [9]. Attempting to improve their scalability to large datasets is a current computational challenge.

Another significant disadvantage of Bayesian methods, at least in their simplest form, is their inability to detect cycles. In other words, the conditional probabilities only flow one way, so they will be unable to detect something like a feedback loop, which is a common motif of GRNs. However, a subtype of Bayesian methods has been developed in an attempt to correct for this problem: Dynamic Bayesian Networks (DBNs). While a node in simple Bayesian methods corresponds to a gene, a node in DBNs corresponds to a gene \textit{at a specific timepoint}. While this solves the problem of detecting cycles (and allows us to detect self-edges), it also presents some new challenges. First of all, either temporal or pseudo-temporal data is required. Obtaining temporal data may not always be feasible, and pseudo-temporal ordering is a complex computational challenge in itself (this will be discussed in the next section). Also, DBNs are more computationally expensive than the already expensive simple Bayesian methods. However, despite these challenges, DBNs are still fairly widely used [21,22,23].

\subsection{Information Theory}

Information theory, first developed by Claude Shannon in 1948, provides a theoretical framework for quantifying information and studying its properties. One fundamental measure of information theory is entropy, which gives a numerical measure of a random variable's ``uncertainty''. For a discrete random variable $X$, the entropy is defined as:

\begin{equation}
H(X)= - \sum_{x \in X}p(x)log(p(x)) 
\end{equation}
where $p(x)$ is the probability distribution of the random variable. If the expression data is continuous, then it must first be discretized through a binning process before being used in this formula, and development of an optimal binning strategy is itself an active area of research [29,30]. 
For two random variables $X_i$ and $X_j$, one can quantify their ``mutual information'' as the amount by which the entropy of their joint distribution is reduced compared to their combined individual entropies:
\begin{equation}
I(X_i,X_j)= \sum_{x_i \in X_i}\sum_{x_j \in X_j}p(x_i,x_j)log\frac{p(x_i,x_j)}{p(x_i)p(x_j)}
\end{equation}
where $p(x_i,x_j)$ is the joint probability distribution of $X_i$ and $X_j$.  

In network inference, mutual information can serve as a measure of dependence between genes. Since it is a symmetric measure, $I(X_i,X_j)$ is assigned as the weights of both edges $X_i\rightarrow X_j$ and $X_j\rightarrow X_i$ (this can also be represented as the undirected edge $X_i\leftrightarrow X_j$). Despite this disadvantage of not predicting causal direction, mutual information has the advantages of being able to detect non-linear interactions (unlike simple correlation measures) and of being scalable to whole-genome networks. In general, information theory-based algorithms tend to perform better than correlation methods, but experience similar biases: increased detection of feed-forward loops, fan-ins, and fan-outs, but also an increased rate of false positives for cascades [9].

One of the most famous network inference methods based on information theory is ARACNE [25], introduced by Margolin et al. in 2006. In this method, the mutual information is estimated for each pair of genes, and then assigned as their edge weight. Then, all edges with a weight below a certain threshold of statistical significance are eliminated. Finally, and perhaps most importantly, the remaining edges are pruned according to the Data Processing Inequality. For each triplet of nodes $X_i$, $X_j$, and $X_k$, the following inequality is checked:

\begin{equation}
I(X_i,X_k)\leq \min \{ I(X_i,X_j),I(X_j,X_k) \}
\end{equation}

If this statement is true, then the edge $X_i\leftrightarrow X_k$ is eliminated. The goal of this pruning step is to yield a sparse network with minimal redundancy and maximal explanatory power. This is only a brief overview of ARACNE, and for a more in-depth explanation we refer readers to the original paper [25]. 

Other methods based on information theory and similar to ARACNE include Butte and Kohane's method [24], CLR [26], and MRNET [27]. Several newer algorithms also make use of concepts from information theory. An exciting new method is Partial Information Decomposition and Context (PIDC) [28]. PIDC computes informational relationships in a triplet-wise, rather than pair-wise, manner, so as to determine the proportion of the mutual information between two genes that cannot be explained in terms of any other third gene, thereby eliminating indirect and redundant relationships in the predicted network. For a detailed explanation of how this is calculated, please see the original paper [28].

\subsection{Phixer}

Another new algorithm that is different from but inspired by information theory is Phixer, introduced in Singh et al. 2018 [31]. Phixer has the advantage of producing an output graph that is both directed (unlike most information theory methods) and can contain cycles (unlike most simple Bayesian methods). In Phixer, one computes the so-called $\phi$-mixing coefficient for the edge $X_i\rightarrow X_j$ as

\begin{equation}
\phi (X_j | X_i)= \max_{S\subseteq A, T\subseteq B} | Pr \{ X_j \in S | X_i \in T \}-Pr \{ X_j \in S  \} |
\end{equation}
which in essence quantifies the maximum distance between conditional probability of $X_j$ given $X_i$, and the unconditional probability of $X_j$. Here $A$ and $B$ are the finite sets in which the random variables $X_j$ and $X_i$ take values, respectively. 
 $S$ and $T$ represent the different subsets of $A$ and $B$, and $\phi (X_j | X_i)$ is the maximum distance between the conditional and the unconditional probability across subsets. 
 
 The $\phi$-mixing coefficient comes with some useful properties. It is bounded in the interval $[0,1]$, and $\phi\left(X_{i}|X_{k}\right)=0$ if and only if $X_i$ and $X_k$ are independent. Moreover, unlike correlation or mutual information it is asymmetric $\phi\left(X_{i}|X_{k}\right)\neq \phi\left(X_{k}|X_{i}\right)$, and hence can discriminate the direction of the influence. 

The inference starts by first computing the $\phi$-mixing coefficient for each directed edge between two nodes, and then the network is pruned to eliminate redundant edges. For every possible triplet of nodes $X_i$, $X_j$, and $X_k$, the following inequality is checked:
\begin{equation}
\phi (X_k | X_i) \leq \min \{ \phi (X_j | X_i),\phi (X_k | X_j) \}
\end{equation}
If this statement is true, then the edge $X_i\rightarrow X_k$ is eliminated, and hence the statistical relationship between $X_i$ and $X_k$ can be fully explained by the edges $X_i\rightarrow X_j$ and $X_j\rightarrow X_k$. This pruning method is similar to the Data Processing Inequality (Equation 4), used in ARACNE [25]. Importantly, the goal of the pruning step is to produce the most parsimonious network consistent with the data, which is not necessarily the same as producing the most accurate network prediction [32]. So, depending on their priorities, researchers may choose to include or omit the pruning step.

\subsection{Miscellaneous}

In addition to correlation, regression, Bayesian networks, information theory, and Phixer, many more interesting and creative types of network inference algorithms exist. Some examples are Gaussian graphical models [33,34], ODE-based methods [35,36], Boolean methods [37,38], and deep learning with neural networks [39]. Unfortunately, we do not have time to summarize every single method here, so we refer the reader to the original papers.

\section{Computational Challenges}

Although the field of network inference has progressed a great deal since its inception, there are several problems that are unsolved and require more research. In this section, we will discuss some of these problems.

\subsection{Combining Algorithms}

It has long been thought that combining an assortment of different algorithms could prove beneficial in network inference. Hill et al. 2016 [40] confirmed this empirically with a computational experiment in which they aggregated results from randomly-chosen inference algorithms, and evaluated the results for accuracy using the area under the ROC curve measure. The results showed a general trend in which the more algorithms were included in the aggregation, the more accurate the results were. Another similar computational experiment was performed in which the results from the top performing algorithms were aggregated (rather than the algorithms being randomly chosen). In this case, the aggregated results generally achieved higher accuracy than even the best individual algorithms. (Please note that Hill et al. 2016 is about the inference of networks from phosphoprotein data, which is computationally very similar to gene network inference.) These results confirm the findings of an earlier analysis found in Marbach et al. 2012 [9]. However, while there is ample evidence that combining algorithms can be beneficial, more research is needed to find the optimal combination strategy.

\subsection{Multiple Data Sources}

Up to this point in the paper, we have been under the assumption that we have only gene expression data to work with when inferring the structure of a GRN. However, in some cases, we may have access to more information. So, an interesting current problem is how best to combine information from multiple data sources in order to generate a prediction.

Some recent papers on the integration of multiple data sources are Yuan et al. 2018 [41], Liang et al. 2019 [42], Lam et al. 2016 [43], and Aibar et al. 2017 [44]. Yuan et al. 2019 combines gene expression data with DNA methylation and copy number variation data. Liang et al. 2019 combines gene expression data with genome-wide binding data, gene ontologies, pathway data, and ChIP-Seq data. Lam et al. 2016 is an especially creative application of this concept. Here, analysis of gene expression data from the species in question is combined with analysis of gene expression data from homologous species. Aibar et al. 2017 introduces SCENIC, a method that combines the raw results of a GENIE3 [16] analysis with transcription factor binding motif information from RcisTarget (also introduced in [44]) to select out a subset of high-confidence interactions, and then uses AUCell (also introduced in [44]) to classify the cells into transcriptional states.

In some cases, it may be necessary to study a change in a GRN, for which something is already known about the GRN's prior structure. This could be in the context of cellular differentiation, or in a pathological context, as with cancer. In these cases, the challenge is to incorporate the prior structural knowledge into the inference of the new structure. Some algorithms have been developed specifically for this purpose [33,45]. Also of note is the database Transcriptional Regulatory Relationships Unravelled by Sentence-based Text-mining (TRRUST) [46], which contains known gene regulatory interactions in humans and mice. This resource can be used for constructing prior networks from which to infer the reprogrammed differential networks [33].

\subsection{Pseudo-Temporal Ordering}

As discussed before, most of the expression data we rely on for network inference is static data collected by methods such as RNA-Seq. This is disadvantageous, in part, because it doesn't allow us to detect self-edges. In an ideal situation, we would have dynamic data for each gene. While this is not currently feasible, some methods have been developed which attempt to create ``pseudo-time series'' data from static data [47,48,49,50]. An excellent review of pseudo-temporal ordering methods is Cannoodt et al. 2016 [51].

Sanchez-Castillo et al. 2017 [52] is a great example of how pseudo-temporal ordering can be can be useful in network inference. The researchers attempt to infer the GRN structure of a set of 48 genes, using a sample of 442 expression profiles from mouse embryos. The expression profiles are static, from single-cell qPCR, but the algorithm the researchers are using requires time series data. So, they employ the MOLO algorithm [53] to construct a pseudo-time series. The researchers then infer the GRN and draw biological insights about cell differentiation in mice from its structure. Moreover, a computational experiment is conducted to show how differences in the temporal order can affect the results of the network inference. In addition to the analysis of the mouse embryo data, another inference analysis is performed on pseudo-temporally ordered RNA-Seq data from zebrafish.

However, while pseudo-temporal ordering has been useful in some cases, it has also faced criticism. Moris et al. 2016 [54] questions one of the underlying assumptions of many of these algorithms, that cell fate transitions are smooth and continuous. More research is needed on the subject to improve upon these algorithms and address these criticisms.

It should also be noted that most of the work on pseudo-temporal ordering has focused on cellular differentiation, and the application of these methods to cells that switch between transient expression states is a more difficult problem that requires further study.

\section{Applications in Cancer Research}

While network inference can be an invaluable tool for the investigation of many different diseases, the most obvious applications are in cancer research. Some important roles of GRNs are to regulate cell cycle timing, proliferation, and apoptosis. Cancer arises from a loss of regulation of these processes, and understanding the structure of the underlying networks can yield powerful insights into discovery of relevant genes [55], clinical outcome prediction [56], drug target identification [57,58,37], elucidation of sex-linked differences [59], investigation of transcriptional reprogramming [33], and more. For example, Figure 4 shows as inferred network of genes using Phixer from [60] that is critical in driving drug resistance in BRAF V600E-mutated melanoma. Also, many of the papers previously cited include a testing section in which the algorithm is applied to a cancer dataset to test its efficacy [14,21,41,42,33,34,37,31,44].

   \begin{figure}[thpb]

      \centering

      \includegraphics[scale=.45]{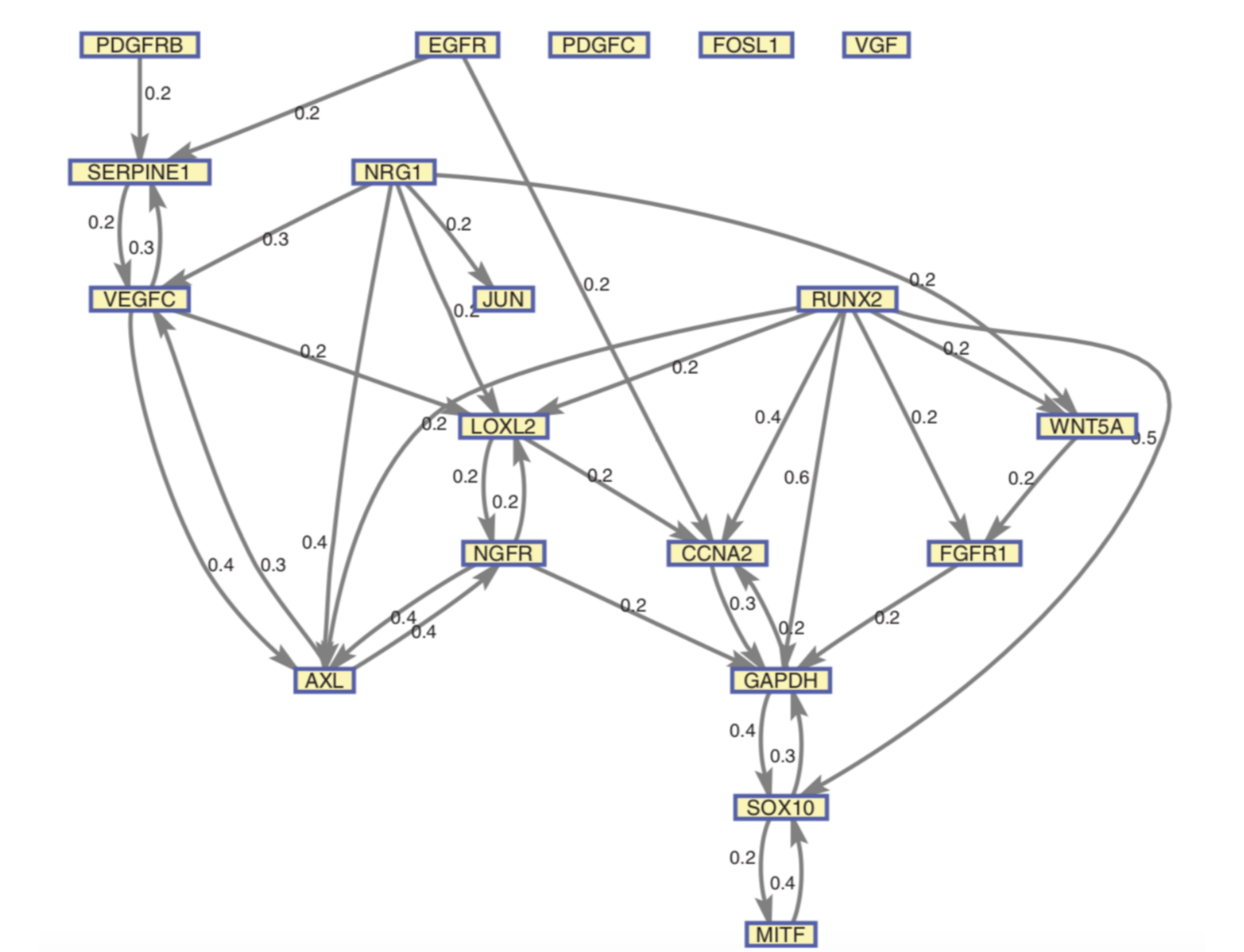}
      \caption{Gene regulatory network inferred in [60] in the context of drug resistance in melanoma. More specifically, RNA Fluorescent In Situ Hybridization (FISH) was used to count the number of mRNAs of $19$ different genes in single melanoma cells. The joint distribution of mRNA levels measured across thousands of cells was then used to infer the network using Phixer. This network was shown to play a key role in driving drug-sensitive cells into a transient drug-tolerant state even in the absence of the drug. Such drug-tolerant cells survive exposure to the drug, and drug-induced reprogramming of this network allows these to become stably resistant to the drug.}
      \label{figurelabel}
   \end{figure}

There have been so many exciting new applications of network inference that we cannot provide an in-depth explanation of all of them here. So, we will focus on one specific paper to serve as a case-study: Moore et al. 2019 [61]. This is an excellent example of how network inference can be used to understand cancer at the cell systems level. The researchers apply the BC3Net [62] inference algorithm to the gene expression profiles of 333 prostate cancer patients, obtained from The Cancer Genome Atlas (TCGA) [63]. The resulting network is then analyzed, using Gene Pair Enrichment Analysis (GPEA) [64] and the Cancer Gene Census [65], with a focus on gene interactions that the researchers feel could be exploited for clinical benefits, including targeted therapy. This paper is valuable not only because of its interesting insights on prostate cancer, but also because the methodology is described in such a clear way that it could serve as a step-by-step guide for researchers hoping to conduct their own analysis of data from another disease. Similar analyses have been performed on data from lymphoma [66], colon cancer [67], and breast cancer [68], but there are still hundreds of different cancer types and subtypes that have yet to be analyzed, many of which are available on TCGA and just waiting for an eager computational biologist to take up the challenge.

\section{Tools}

Here are some useful, freely available tools for network inference researchers:

\begin{itemize}
\item DREAM Challenges [8] - network inference competitions, for which gold standard datasets are made publicly available. These are often used for benchmarking. Available online at: http://dreamchallenges.org

\item The Cancer Genome Atlas (TCGA) [63] - a collection of genomic, epigenomic, transcriptomic, and proteomic cancer data, including expression profiles that can be used for gene network inference. Available online at: https://portal.gdc.cancer.gov

\item TRRUST [46] - a database of experimentally-validated gene interactions in humans and mice. Available online at: https://www.grnpedia.org/trrust

\item Cancer Gene Census [65] - a catalogue of genes known to be related to cancer. Available online at: https://cancer.sanger.ac.uk/census

\item The Cancer Network Galaxy (TCNG) - an online database of cancer gene networks, predicted from public expression data. Not yet officially published, but a beta release is available online at: http://tcng.hgc.jp

\item GeNeCK [69] - an online tool for gene network inference and visualization, for which the user can choose from 8 different inference algorithms. Available online at: http://lce.biohpc.swmed.edu/geneck
\end{itemize}

\section{Conclusion}

In this paper, we have given a brief overview of the basics of network inference, different types of algorithms, some unsolved computational challenges, and exciting new applications in cancer research. While we hope this review has been helpful, it is not completely comprehensive, and we encourage the reader to further explore this topic by reading the papers cited here, and by testing out the algorithms and data sources for themselves.

\section*{References}

\bibliography{mybibfile}

\end{document}